\documentclass[10pt,twocolumn,letterpaper]{article}

\usepackage{cvpr}              

\usepackage{graphicx}
\usepackage{amsmath}
\usepackage{amssymb}
\usepackage{booktabs}
\usepackage{bm}
\usepackage{dsfont}
\usepackage{subcaption}
\usepackage{comment}
\usepackage[accsupp]{axessibility}

\usepackage[pagebackref,breaklinks,colorlinks]{hyperref}

\usepackage[capitalize]{cleveref}
\crefname{section}{Sec.}{Secs.}
\Crefname{section}{Section}{Sections}
\Crefname{table}{Table}{Tables}
\crefname{table}{Tab.}{Tabs.}

\def\cvprPaperID{3880} 
\def\confName{CVPR}
\def\confYear{2023}

\makeatletter
\def\thanks#1{\protected@xdef\@thanks{\@thanks
        \protect\footnotetext{#1}}}
\makeatother

\begin{document}

\title{ Weakly supervised segmentation with point annotations for histopathology images  via contrast-based variational model}

\author{Hongrun Zhang$^{1,4\dag}$, Liam Burrows$^{2\dag}$, Yanda Meng$^{1}$, Declan Sculthorpe$^{3}$,  Abhik Mukherjee$^{3}$, \\ Sarah E Coupland$^{4}$, Ke Chen$^{2}$, Yalin Zheng$^{1\ast}$ \\
\thanks{\dag: Equal contribution; $\ast$: Corresponding author}
$^1$Department of Eye and Vision Science, University of Liverpool, Liverpool, UK
$^2$ Department of Mathematical \\ Sciences and Centre for Mathematical Imaging Techniques, University of Liverpool, Liverpool, UK\\
$^3$ Biodiscovery Institute, School of Medicine, University of Nottingham, Nottingham, UK \\
$^4$ Institute of Systems, Molecular and Integrative Biology, University of Liverpool, Liverpool, UK\\
{\tt\small \{hongrun.zhang,liam.burrows,yanda.meng,s.e.coupland,k.chen,yalin.zheng\}@liverpool.ac.uk } \\
{\tt\small \{declan.sculthorpe,abhik.mukherjee1\}@nottingham.ac.uk, zhang.hr.jlu@gmail.com} \\
}
\maketitle

\begin{abstract}

Image segmentation is a fundamental task in the field of imaging and vision. Supervised deep learning for segmentation has achieved unparalleled success when sufficient training data with annotated labels are available. However, annotation is known to be expensive to obtain, especially for histopathology images where the target regions are usually with high morphology variations and irregular shapes. Thus, weakly supervised learning with sparse annotations of points is promising to reduce the annotation workload.  In this work, we propose a contrast-based variational model to generate segmentation results, which serve as reliable complementary supervision to train a deep segmentation model for histopathology images. The proposed method considers the common characteristics of target regions in histopathology images and can be trained in an end-to-end manner.  It can generate more regionally consistent and smoother boundary segmentation, and is more robust to unlabeled `novel' regions. Experiments on two different histology datasets demonstrate its effectiveness and efficiency in comparison to previous models. Code is available at: \url{https://github.com/hrzhang1123/CVM_WS_Segmentation}.

\end{abstract}


\section{Introduction}

\begin{figure}[t]
  \centering
   \includegraphics[width=0.8\linewidth]{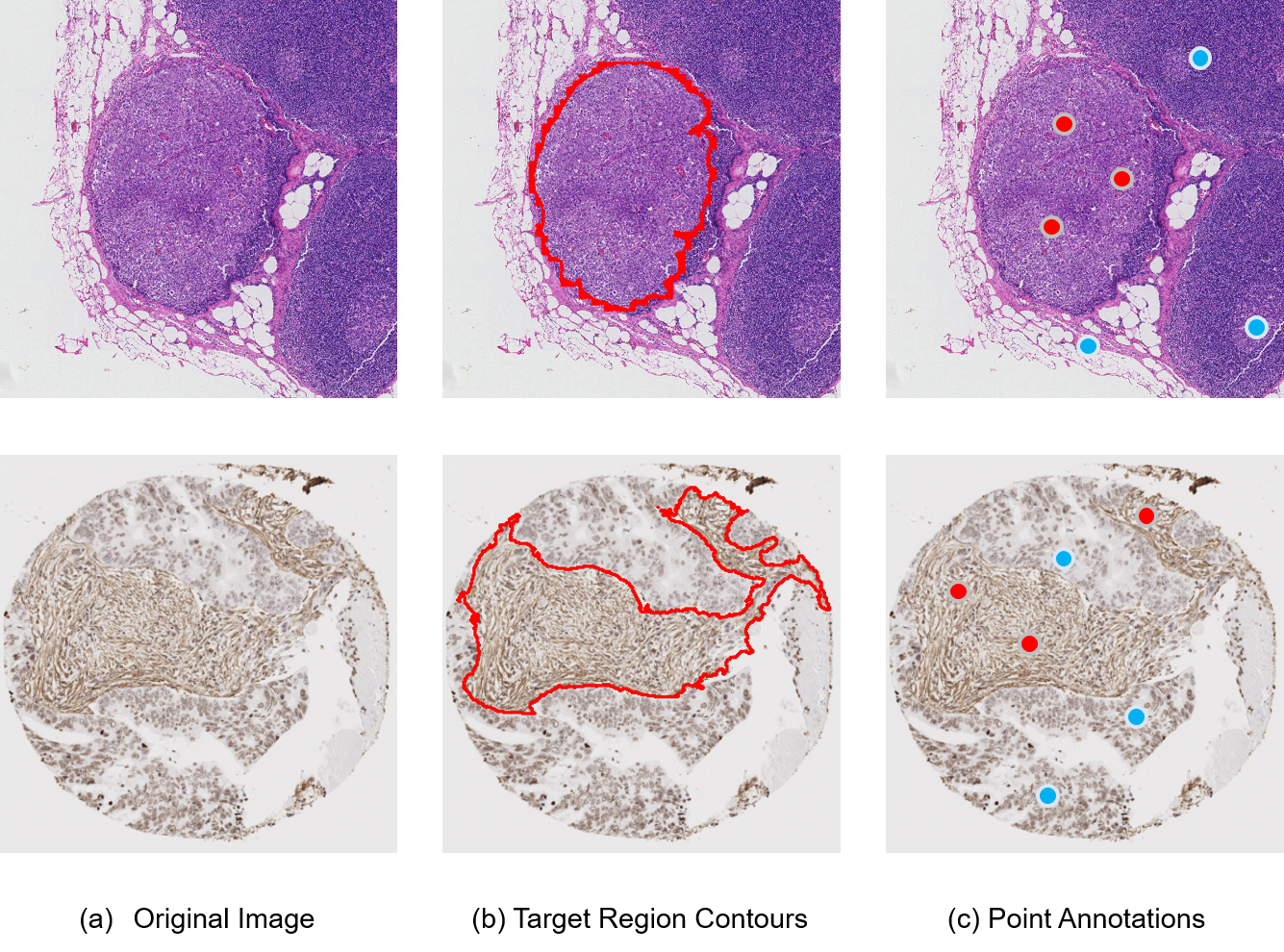}
   \caption{Two examples of histopathology images with target regions (e.g. tumor (top row) and stroma (bottom row)) annotated by contours (b) or in-target  and out-of-target points (c). }
   \label{fig_annotation}
\end{figure}

Histopathology images are of great importance for clinical diagnosis and prognosis of diseases. With the thriving of artificial intelligence (AI) techniques over the past decade, especially deep learning, automatic analysis of histopathology images in some tasks has achieved comparable or even surpassing performance in comparison with human pathologists' reviewing \cite{litjens2016deep,zhang2020piloting, coupland2023application, bejnordi2017diagnostic}. However, most competent methods are based on supervised learning, and their performances critically rely on a large number of training samples with detailed annotations. Yet such annotations usually require experienced pathologists and are expensive (in terms of cost and time consumption) to obtain, and also subject to human errors. The annotation problem for histopathology images is particularly demanding, not only due to the large size of such an image but also resulting from irregular shapes of target tissues to be annotated (See Figure.\ref{fig_annotation}). 

Weakly supervised learning is a promising solution to alleviate the issues of obtaining annotations. The annotations for the ``weakly supervised" can specifically refer to image-level labelling (multiple instance learning) \cite{li2021dual,shao2021transmil,zhang2022dtfd,li2023task,lin2023interventional,xiang2023exploring}, partial annotations within an image (point or scribble annotation) \cite{lin2016scribblesup}, or full annotation in partial images (semi-supervised learning) \cite{luo2021semi}. Amongst these three categories, learning by partial annotations has excellent target localization capability, yet requires comparably less cost to annotate. 

Interactive segmentation with scribbles or points has been widely studied for a few decades \cite{boykov2001interactive}. Conventional methods relied on user interactive input for object segmentation, such as grab-cut \cite{rother2004grabcut}, Graph-cut \cite{boykov2001interactive}, active contour-based selective segmentation \cite{roberts2019convex}, random walker \cite{grady2006random}. In recent years it has been a hot topic to develop segmentation models that can be trained by utilizing only the scribble or point annotations, formulated as partially-supervised learning problem \cite{lin2016scribblesup,zhao2020weakly, peng2019semi, valvano2021learning, li2019weakly,yoo2019pseudoedgenet,qu2020weakly, lee2020scribble2label, can2018learning, cheng2020self}. Yet, existing partially-supervised methods were designed mainly for natural images or medical images with relatively regular-shaped objects and consistent textures, and very few are directly applicable to histopathology images given the above challenges.

In this work, we focus on partially-supervised learning for histopathology image segmentation based on in-target and out-of-target point annotations, where in-target points refer to those labelled inside the target regions and out-of-target points refer to the outside ones. Histopathology images are significantly distinguished from other types of images. In many histopathology images, the target objects present distinct regional characteristics. Specifically, as the example shown in Figure.\ref{fig_annotation}, the tumors usually cluster inside large regions, in which morphological features or textures are similar and visually different to the outside regions, and there exist comparably clear boundaries. Existing works on partially-supervised learning do not well utilize this characteristic. Moreover, a histopathology image scanned from a tissue section often contains some non-target regions that are visually or morphologically unique to those in other images. If such regions are not labelled, they will be `novel' to a trained machine learning model, and the predicted categories of them will depend on their similarity to the neutral tissue and to the target tissue. If such novel regions are more similar to target tissues, they will be wrongly predicted as the target category, leading to false detections. Existing methods are limited in tackling this situation, especially for those methods based on consistency training on data augmentation \cite{liu2022weakly, gao2020renal}, as consistency supervision may amplify such errors. 

Based on the above observations and insights, we propose to adopt a variational segmentation model to provide complementary supervision information to assist the training of a deep segmentation model. This variational segmentation model itself will be guided by annotated in-target and out-of-target points for the segmentation of target regions in the images. Variational methods are powerful tools for segmenting structures in an image. Often posed as an optimisation problem, energy functionals can be carefully constructed to satisfy certain desired properties, for example, maintaining consistency inside the evolving contour, or constraining the length of the boundary to ensure a smooth boundary \cite{chan2001active,rudin1992nonlinear}.

The uniqueness of variational methods highly fits the characteristic of target regions in histopathology images, as mentioned above. However, existing variational methods cannot be directly applied to high-dimensional deep features, which contain higher-level semantic information. To tackle this problem, we introduce the concept of contrast maps derived from deep feature correlation maps and formulate a variational model applied to the obtained contrast maps.  Specifically, a set of correlation maps are generated based on the annotated points on an image. The corresponding contrast maps can then be obtained by the pairs of correlation maps of in-target and out-of-target points through the subtraction operation. A variational formulation is used to aggregate the obtained contrast maps for the final segmentation result. Finally, the variational segmentation provides complementary supervision to the deep segmentation model through the uncertainty-aware Kullback–Leibler (KL) divergence. Besides, the proposed model can alleviate the aforementioned issue of unlabeled novel tissue regions, resulting from the subtraction operation in obtaining the contrast maps. 

In summary, the main contribution of this work is the formulated contrast-based variational model, used as reliable complementary supervision for training a deep segmentation model from weak point annotations for histopathology images. The variational model is based on the proposed new contrast maps, which incorporate the correlations between each location in an image and the annotated in-target and out-of-target points. The proposed model is well suited for the segmentation of histopathology images and is robust to unlabeled novel regions. The effectiveness and efficiency of the proposed method are empirically proven by the experimental results on two different histology datasets.

\begin{figure}[h]
     \centering
     \begin{subfigure}[b]{0.5\textwidth}
         \centering
         \includegraphics[width=1\linewidth]{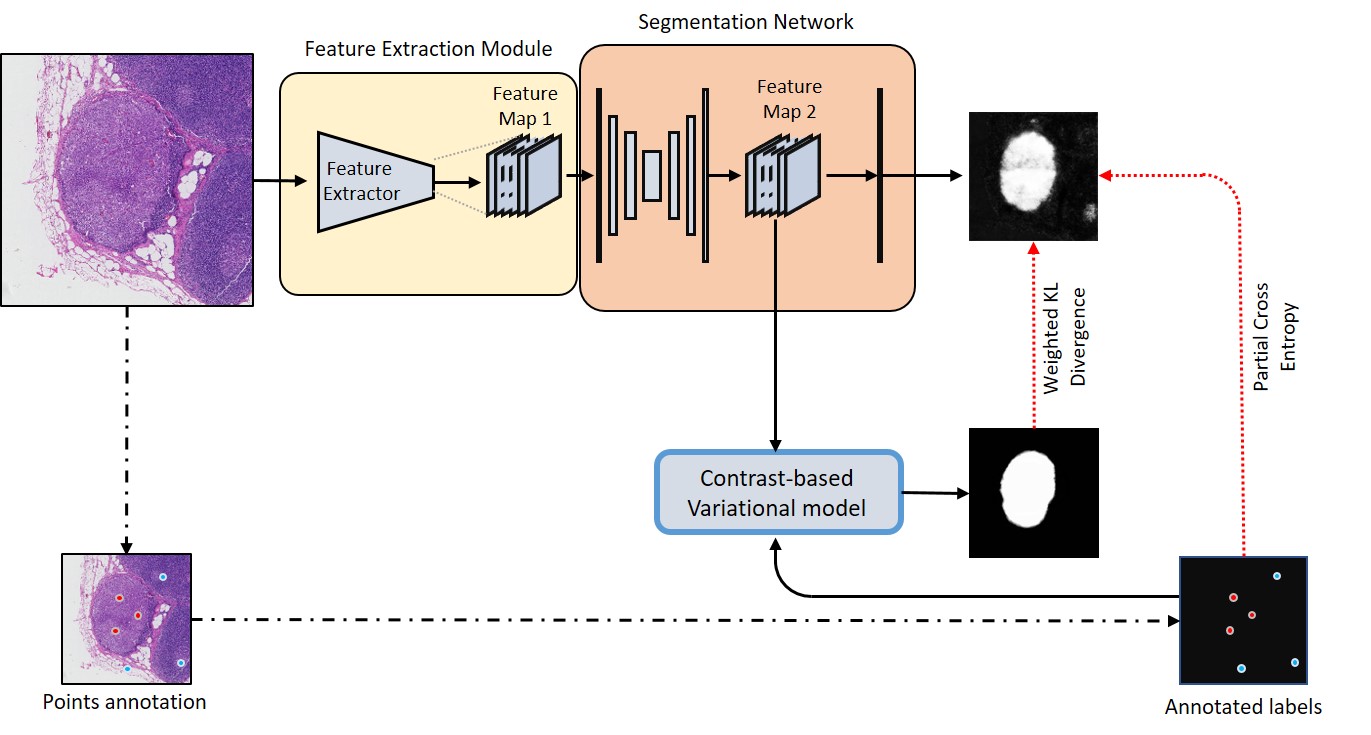}
         \caption{ Overview of the end-to-end training process. An image with a high resolution is fed into the deep segmentation model (comprised of a feature extractor and a segmentation network) which outputs the segmentation result. Meanwhile, the contrast-based variational model is fed with the feature map from the convolutional segmentation network and the annotated points and generates the variational segmentation. The deep segmentation model is supervised by the variational segmentation and the annotated points simultaneously by the weighted KL divergence and the partial cross-entropy, respectively. The whole network can be trained in an end-to-end manner. In particular, the variational model is only used to train the deep segmentation model and is not used in the inference stage. The red dots represent the annotated in-target points, while the blue dots are the annotated out-of-target points. }
         \label{overview}
     \end{subfigure}
     \hfill
     \begin{subfigure}[b]{0.5\textwidth}
         \centering
         \includegraphics[width=1\linewidth]{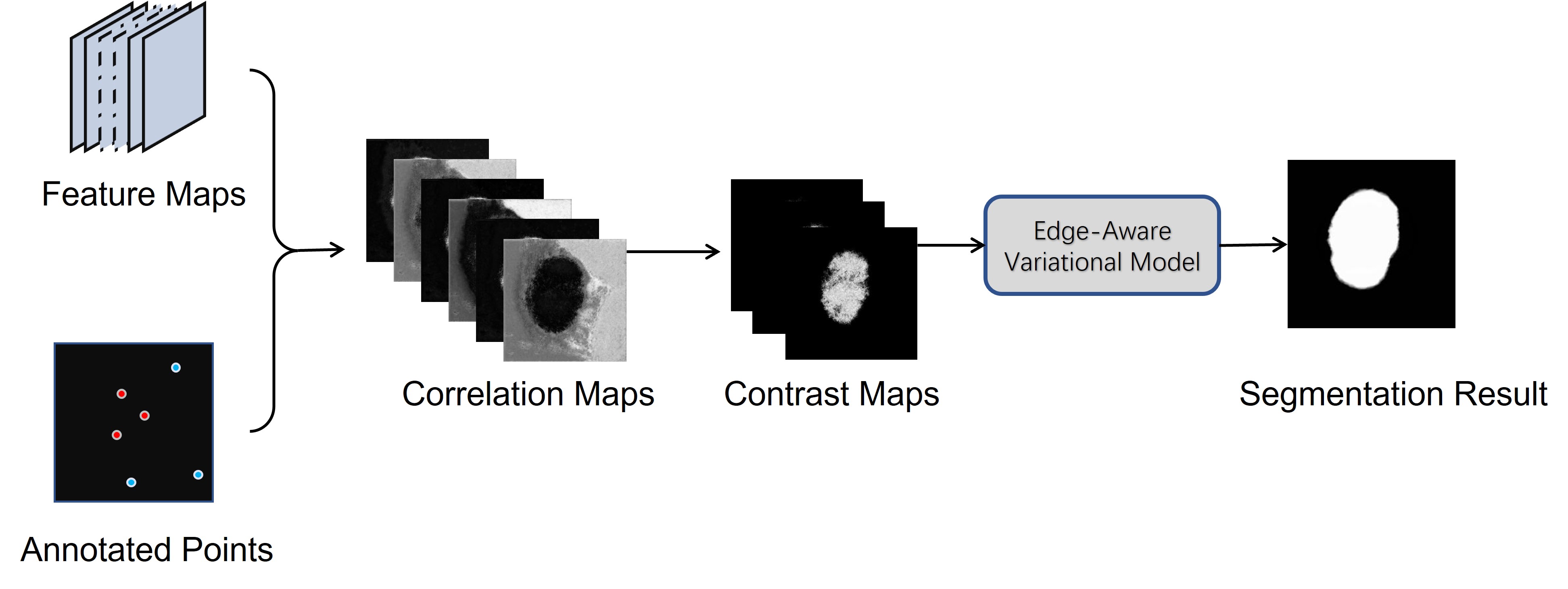}
         \caption{Diagram of the proposed contrast-based variational model.}
         \label{sfig_contrast_varational}
     \end{subfigure}
        \caption{Overview of the proposed method.}
        \label{fig_ablation}
\end{figure}

\section{Related works}
\subsection{Variational segmentation methods} 
Variational segmentation methods have been well studied over the past few decades. The seminal work by Mumford and Shah \cite{mumford1989optimal} is perhaps the most influential variational model, aiming to find a piecewise smooth approximation to a given image, as well as a discontinuity set of edges. As it is difficult to solve directly many works have their roots in the Mumford-Shah model, with variants proposed \cite{ambrosio1990approximation,chambolle1999finite} and convex relaxations \cite{cai2013two,pock2009algorithm}. A famous example is the model by Chan and Vese \cite{chan2001active}, in which the piecewise-smooth condition is relaxed to be piecewise-constant, drastically simplifying implementation. 

Selective segmentation is one type of variational segmentation method that utilizes user input points to segment region of interest. It is usually implemented by incorporating the distance constraints into the energy functional to optimize. The distance constraints are based on the locations of the input points. Some examples are point-wise Euclidean distance constraints \cite{gout2005segmentation,liu2018weighted}, normalised Euclidean distance constraints \cite{spencer2015convex} and normalised edge weighted geodesic distance constraints \cite{roberts2019convex,burrows2020reproducible}. However, conventional selective segmentation methods mainly rely on geometrical relationships and image pixel values, without considering high-level region correlation in an image. Thus they are deficient in dealing with regions of complicated textures and irregular shapes, particularly those in  histopathology images.

\subsection{Partially supervised segmentation}

Most partially-supervised methods for segmentation are based on region or scribble annotations. Point annotation essentially is a special form of scribble annotation, but it provides sparser and weaker supervision information \cite{qu2020weakly, yoo2019pseudoedgenet, berthelot2019mixmatch, li2019weakly, gao2020renal, zhao2020weakly}, in return to speed up the annotation process. 

A large body of methods addresses the partially-supervised learning problem through synthesizing labels generated from partial annotations, by which the learning process virtually becomes fully supervised.  The synthesizing labels can be generated from graphical models \cite{can2018learning, lin2016scribblesup}, auxiliary architectures \cite{liu2022weakly, luo2022scribble}, temporal predictions \cite{lee2020scribble2label}, Voronoi partition \cite{qu2020weakly}, concentric circle extension \cite{li2019weakly} etc. Nevertheless, many existing synthesizing methods are based on the consistency among different augmented versions of the same image \cite{gao2020renal, liu2022weakly, lee2020scribble2label}, which is deficient in dealing with unlabeled novel regions. Particularly, histopathology images usually have some special regions in each that are unique to those in other images.  

Apart from synthesizing labels, segmentation regularization from graphical models has also been proven to be an effective strategy, such as the CRF loss \cite{tang2018regularized}, edge detection \cite{yoo2019pseudoedgenet}, Normal-cut \cite{tang2018normalized}, size constraint \cite{kervadec2019constrained}, level-set \cite{kim2019mumford}. However, these regularizations are less applicable to histopathology images, as the tissue regions usually contain complicated textures or irregular shapes.

\section{Preliminary: Selective segmentation by variational model}
The two-phase segmentation model by Chan and Vese \cite{chan2001active} is widely used and successful, and many subsequent works use the Chan-Vese framework. Of particular interest is the work by Chan \etal \cite{chan2006algorithms}, who reformulated the non-convex Chan-Vese model into a convex version. This was investigated further by Bresson \etal \cite{bresson2007fast}, who proposed incorporating an edge detector into the total variation term. Given an image $f$ defined on a continuous domain $\Omega \subset \mathbb{R}^2$, their model aims to partition it into two phases, $\Omega_1$ and $\Omega_2$, with a set of edges $\Gamma$ defining the boundary, so that $\Omega = \Omega_1 \bigcup \Omega_2 \bigcup \Gamma$. Their model is given as:
\begin{align} \label{eq_edge_cv}
    F_{CCV}(u,c_1,c_2) & = \int_{\Omega}  g(|\nabla f |) \nabla u(\mathbf{x})| \, d\mathbf{x} \\
    & + \int_{\Omega} \big( (f-c_1)^2 - (f-c_2)^2 \big) u(\mathbf{x}) \, d\mathbf{x}, \nonumber
\end{align}
where $c_1$ and $c_2$ are the mean intensities of $f$ inside $\Omega_1$ and $\Omega_2$ respectively, $u(x) \in [0,1]$ and $g(s) = \frac{1}{1+\iota s^2}$ is an edge detector. The presence of $g$ encourages the contour to settle on edges. The region $\Omega_1$ is found by thresholding $u$: $\Omega_1 = \{ \mathbf{x} \in \Omega : u(\mathbf{x}) > \gamma\}$, where $\gamma = 0.5$ is usually fixed.

Regarding selective segmentation models, the convex models proposed by Spencer and Chen \cite{spencer2015convex} and Roberts \etal \cite{roberts2019convex} both take a general form:
\begin{align*}
    F_{SS}(u,c_1,c_2) & = \int_{\Omega} g(|\nabla f|) | \nabla u (\mathbf{x}) | \, d\mathbf{x} + \theta \int_{\Omega} \mathcal{D} u (\mathbf{x}) \, d \mathbf{x} \nonumber \\
    & \, + \lambda \int_{\Omega} \big( (f - c_1)^2 - (f-c_2)^2 \big) u(\mathbf{x})  \, d\mathbf{x}.
\end{align*}
The addition of the distance constraint, $\mathcal{D}$, forces the region of interest to be close to the annotated points. The distance constraint in the Spencer-Chen model is a normalised Euclidean distance, whereas in Roberts model is a normalised edge weighted geodesic distance. See \cite{spencer2015convex,roberts2019convex} for details on definitions for these terms.

\section{Method}

\subsection{Network architecture}

Histopathology images usually have high resolutions. Direct segmentation on lower-resolution (down-sampled) images may lose relevant information contained in higher-resolution details. To preserve such information, as well as to fit a histopathology image into a deep neural network with the consideration of memory and computation limits of a computing unit (typically a GPU), we adopt a two-component architecture as shown in Figure.\ref{overview}, which enables the end-to-end training and inference. The first component is a feature extractor (the yellow box in Figure.\ref{overview}). When fed with a high-resolution image, the feature extractor will generate a reduced-dimension feature map that extracts the high-level representation of the input image. The generated feature map is then forwarded into the second component, a regular deep segmentation network for segmentation (the orange box in Figure.\ref{overview}). The segmentation network then outputs segmentation results.  Images of regular sizes can skip the feature extraction module and be fed directly to the segmentation network. The feature extractor and the segmentation network combined is defined as the deep segmentation model.

The feature extractor can either be a convolution network or a vision transformer \cite{dosovitskiy2020image}. Without loss of generality, we used the convolutional module in the ResNet34 \cite{he2016deep} as the feature extractor. The U-Net \cite{ronneberger2015u} is adopted as the segmentation network. Note that another feature extraction option is to extract a feature vector from each patch from the image, and stitch all the feature vectors into a feature map \cite{takahama2019multi}. This is the typical operation to process gigapixel whole slide images with deep learning model. Also, note that the optimization of a patch-level feature extraction model can be implemented by multi-stage back-propagation \cite{takahama2019multi}.

\subsection{Loss function for training}

We consider only the binary segmentation for histopathology images, i.e., the target and the non-target regions. The extension of the proposed method to multiple categories will be straightforward. 

Let $\hat{\bm{y}}$ be the segmentation result, i.e., the probability map of being the target regions outputted from the deep segmentation model. In the proposed method, the overall loss function to  train the deep segmentation model is formulated based on two sub-loss functions, namely, the partial cross entropy ($\mathcal{L}_{pce}$) and the weighted KL divergence ($\mathcal{L}_{wkl}$),
\begin{equation}
    \mathcal{L} = \mathcal{L}_{pce}{(\hat{\bm{y}}, \bm{y}_{S})}  + \mathcal{L}_{wkl}(\hat{\bm{y}}, \bm{u})
\end{equation}

\noindent where $\bm{y}_S$ is the ground-truth labels for the pixel set $S$ that are annotated, and $\bm{u}$ is the segmentation result from the proposed contrast-based variational model (CVM).  

Note that the deep segmentation model trained with only the partial cross-entropy is seen as the \textbf{baseline model}.

\subsubsection{Partial cross-entropy training}

Partial cross-entropy is usually adopted for the case when only partial pixels in an image are with labels. Given an image with the set of annotated pixels $S$, the partial cross-entropy is defined as,
\begin{equation}
\mathcal{L}_{pce}{(\hat{\bm{y}}, \bm{y}_S)} = - \frac{1}{|S|} \sum_{ i \in S }  y_i \log \hat{y}_i + (1-y_i) \log (1-\hat{y}_i)
\end{equation}

\noindent where $\hat{y}_i$  is the  probability of pixel $i$ being in the target region, $y_i$ is the corresponding pixel ground-truth label, and $|S|$ is the size of the annotated pixel set.

\subsubsection{Uncertainty-aware KL divergence loss}

The segmentation result $\bm{u}$ represents the probability map obtained from the CVM. It provides complementary supervision information to train the deep segmentation model. To implement, one straight-forwarding way is to cutoff $\bm{u}$ into a binary mask by a threshold (usually 0.5), and then use it as the pseudo ground-truth labels in the cross-entropy for training. However, the pixels in $\bm{u}$ with high uncertainty values may introduce noise labels in this way. Alternatively, we adopt the weighted KL divergence to directly align the distributions of output segmentation results $\hat{\bm{y}}$ and $\bm{u}$. 
\begin{align*}
  & \mathcal{L}_{wkl}(\hat{\bm{y}}, \bm{u}) = \frac{1}{|K|} \sum_{ i \in K} w_i D_{\textrm{KL}}(u_i||\hat{y}_i) \\
  & = - \frac{1}{|K|} \sum_{i \in K} w_i \left( u_i \log{ \frac{u_i}{\hat{y}_i}     } + (1-u_i) \log{ \frac{1-u_i}{1-\hat{y}_i} }  \right)  
\end{align*}

\noindent where $K$ is the pixel index set of the image. The weight $w_i$ is based on entropy measurement, defined as,
\begin{align*}
w_i & = e^{-2h_i} \\
h_i & = -u_i \log u_i - (1-u_i) log (1-u_i)
\end{align*}

\noindent where $h_i$ is the entropy value. A larger entropy value (a probability value around 0.5) indicates a larger uncertainty, leading to a smaller value of $w_i$, thus the corresponding pixel in $\bm{u}$ contributes less in the $\mathcal{L}_{wkl}$.

\subsection{Contrast-based variational model for segmentation}
In this section, the detailed introduction of the proposed CVM is provided, and the diagram is shown in Figure.\ref{sfig_contrast_varational}.

\subsubsection{Contrast Map}
A contrast map is formed based on the corresponding correlation maps of a pair of in-target and out-of-target annotated points.  A correlation map of an annotated point w.r.t. a feature map measures the higher-level correlations between each location in the image and the target location that corresponds to the annotated point.  

Given a feature map $\bm{A} \in \mathds{R}^{C \times W \times H }$ (in this paper it is the Feature Map 2 in Figure.\ref{overview}), and the coordinate of an annotated point $p$ w.r.t. to $\bm{A}$ is $(w,h)$, and $\bm{f}_{w,h} \in \mathds{R}^{C \times 1}$ the feature vector extracted from $\bm{A}$ corresponding to the annotated point, the correlation is defined as the cosine similarity, thus the correlation map of point $p$ w.r.t. $\bm{A}$ is defined as,
\begin{align*}
\bm{\mathcal{S}}_p = \textrm{CSM}(\bm{f}_{w,h}, \bm{A}),
\end{align*}

\noindent where $\bm{\mathcal{S}}_p \in \mathds{R}^{W \times H}$ and the element in it located at $(i,j)$ is, 
\begin{equation}
  \mathcal{S}_{p, (i,j)} = \frac{ \bm{f}_{w,h}^{\textrm{T}}  \bm{a}_{i,j} }{  \| \bm{f}_{w,h} \|_2   \| \bm{a}_{i,j} \|_2   } 
\end{equation}

\noindent where T is the transpose operation, and $\bm{a}_{i,j}$ is the feature vector extracted from $\bm{A}$ located at $(i,j)$. 

Given a pair of in-target and out-of-target annotated points, denoted as $p$ and $q$, respectively, with the corresponding correlation maps, $\bm{\mathcal{S}}_{p}$ and $\bm{\mathcal{S}}_{q}$. The corresponding contrast map $ \bm{\mathcal{C}}_{ p,q} \in \mathds{R}^{W \times H} $  is then defined by the differential operation,
\begin{equation} \label{eq_contrast}
   \bm{\mathcal{C}}_{ p, q } = \mathcal{N}  \big( \left( \textit{ReLU} \left( \bm{\mathcal{S}}_{p} - \eta \bm{\mathcal{S}}_{q}  \right) \right)^2 \big),
\end{equation}

\noindent where $\textit{ReLU}$ is the regular ReLU function, i.e., $\textit{ReLU}(x)=x$ if $x>0$ and $\textit{ReLU}(x)=0$ otherwise. $\mathcal{N}$ is the 2-dimensional normalization operation, that will transform the values of the elements in a 2-dimensional matrix (or map) to be between $0$ and $1$. $\eta \in [ 0,1]$ is a positive constant. The subtraction operation will reduce the activation not only from the non-target regions, but also from the unlabelled novel regions, as they probably will have similar activation in both $ \bm{\mathcal{S}}_{p}$ and $\bm{\mathcal{S}}_{q}$ due to high uncertainty.

\subsubsection{Contrast-based variational model}

Let $P$ and $Q$ denote the sets of in-target and out-of-target annotated points, respectively, and the corresponding set of contrast maps $\bm{\mathcal{C}}_{p,q}, \ p \in P, \ q \in Q $. For each in-target point $p$, an edge-aware variational CV energy function is optimized w.r.t. to the mean map of the contrast maps regarding $p$,
\begin{align}
\label{eq:contrastive_variationalmodel}
    & F(u_p,c_{1,p},c_{2,p})  = \int_{\Omega} g(|\nabla z_p(\mathbf{x}) |) | \nabla u_p (\mathbf{x}) | \, d\mathbf{x} \\
    & \, + \lambda \int_{\Omega} \big( (z_p(\mathbf{x}) - c_{1,p})^2 - (z_p(\mathbf{x}) -c_{2,p})^2 \big) u_p (\mathbf{x}) \, d\mathbf{x} \nonumber ,
\end{align}
where $z_p(\mathbf{x}) = \frac{\sum_{q} \mathcal{C}_{p,q}(\mathbf{x}) }{|Q|}$. 

Optimisation is done alternatively between the three variables. Minimisation with respect to $c_{1,p}$ and $c_{2,p}$ give closed-form solutions given as:
\begin{align*}
    c_{1,p} = \frac{ \int_{\Omega} z_p(\mathbf{x}) u_p(\mathbf{x}) \, d\mathbf{x} }{\int_{\Omega} u_p(\mathbf{x}) \, d\mathbf{x}}, \ 
    c_{2,p} = \frac{ \int_{\Omega} z_p(\mathbf{x}) (1-u_p(\mathbf{x})) \, d\mathbf{x} }{\int_{\Omega} 1-u_p(\mathbf{x}) \, d\mathbf{x}}.
\end{align*}
Minimising with respect to $u_p$ involves finding the Euler-Lagrange equation and using a gradient descent scheme to solve the resulting non-linear PDE:
\begin{align*}
    \frac{\partial u_p}{\partial t} = \nabla \cdot \Big( g \frac{\nabla u_p}{|\nabla u_p|} \Big) - \lambda \big( (z_p - c_{1,p})^2 - (z_p - c_{2,p})^2 \big),
\end{align*}
Efficient numerical schemes can be implemented to solve this PDE. We use an additive operator splitting scheme \cite{weickert1998efficient}, of which more details can be found in the supplementary information.
The final variational segmentation result comes from the averaging of all variational segmentations,
\begin{equation}
    \bm{u}= \mathcal{N} \Big( \frac{ \sum_{p} \bm{u}_p}{|P|} \Big) 
\end{equation}

Note that in the actual training process, $\bm{u}$ is updated and saved for each image only when a new lowest value of partial cross-entropy on the validation set is achieved.

\begin{table*}[]
\caption{The performance comparison of different methods. The best metric values except the fully-supervised are written in bold.}
\label{tab_compare_otherMethod}
\centering
\begin{tabular}{l|cccc}
\hline
                                  & \multicolumn{4}{c}{Camelyon-16}                                      \\ \cline{2-5} 
Methods                           & Dice Coefficient & Accuracy       & Cohen's Kappa  & AUC            \\ \hline
Partial CE \cite{lin2016scribblesup} (Baseline)             & 0.563$\pm$0.032   & 0.784$\pm$0.028 & 0.398$\pm$0.017 & 0.877$\pm$0.019 \\
MixMatch  \cite{gao2020renal}                        & 0.611$\pm$0.045   & 0.816$\pm$0.070 & 0.449$\pm$0.105 & 0.892$\pm$0.049 \\
Uncertain-aware Mean Teacher  \cite{liu2022weakly}    & 0.590$\pm$0.072   & 0.802$\pm$0.071 & 0.424$\pm$0.091 & 0.900$\pm$0.025 \\
CRF Regularization    \cite{tang2018regularized}            & 0.631$\pm$0.096   & 0.838$\pm$0.069 & 0.500$\pm$0.109 & 0.913$\pm$0.043 \\
Dual Branch      \cite{luo2022scribble}                 & 0.631$\pm$0.087   & 0.886$\pm$0.026 & 0.545$\pm$0.087 & 0.926$\pm$0.013 \\
 Peng \etal   \cite{peng2019semi}                              & 0.615$\pm$0.091   & 0.813$\pm$0.079 & 0.472$\pm$0.126 & 0.901$\pm$0.058 \\
Full Supervision                  & 0.802$\pm$0.029   & 0.934$\pm$0.008 & 0.742$\pm$0.034 & 0.978$\pm$0.005 \\ \cline{1-1}
Direct Variational Regularization & 0.603$\pm$0.105   & 0.815$\pm$0.085 & 0.466$\pm$0.135 & 0.891$\pm$0.041 \\
By Contrast Map                   & 0.709$\pm$0.047   & 0.901$\pm$0.011 & 0.626$\pm$0.049 & \textbf{0.948$\pm$0.017} \\
The Proposed method               & \textbf{0.735$\pm$0.027}   & \textbf{0.918$\pm$0.006} & \textbf{0.657$\pm$0.087} & 0.944$\pm$0.006 \\ \hline
                                  & \multicolumn{4}{c}{Colorectal tissue cores}                            \\ \cline{2-5} 
                                  & Dice Coefficient & Accuracy       & Cohen's Kappa  & AUC            \\ \hline
Partial CE \cite{lin2016scribblesup} (Baseline)             & 0.631$\pm$0.129   & 0.838$\pm$0.141 & 0.501$\pm$0.199 & 0.913$\pm$0.031 \\
MixMatch   \cite{gao2020renal}                       & 0.616$\pm$0.041   & 0.757$\pm$0.055 & 0.446$\pm$0.081 & 0.883$\pm$0.012 \\
Uncertain-aware Mean Teacher  \cite{liu2022weakly}    & 0.580$\pm$0.028   & 0.726$\pm$0.058 & 0.396$\pm$0.075 & 0.866$\pm$0.039 \\
CRF Regularization      \cite{tang2018regularized}          & 0.636$\pm$0.013   & 0.815$\pm$0.018 & 0.512$\pm$0.037 & 0.889$\pm$0.011 \\
Dual Branch        \cite{luo2022scribble}               & 0.644$\pm$0.081   & 0.863$\pm$0.026 & 0.563$\pm$0.079 & 0.915$\pm$0.007 \\
   Peng \etal   \cite{peng2019semi}                     & 0.528$\pm$0.097   & 0.624$\pm$0.137 & 0.285$\pm$0.174 & 0.740$\pm$0.181 \\
Full Supervision                  & 0.772$\pm$0.044   & 0.893$\pm$0.013 & 0.699$\pm$0.049 & 0.951$\pm$0.009 \\ \cline{1-1}
Direct Variational Regularization & 0.632$\pm$0.115   & 0.754$\pm$0.181 & 0.468$\pm$0.236 & 0.888$\pm$0.076 \\
By Contrast Map                   & 0.678$\pm$0.050   & 0.871$\pm$0.029 & 0.598$\pm$0.057 & 0.920$\pm$0.013 \\
The Proposed method               & \textbf{0.710$\pm$0.025}   & \textbf{0.872$\pm$0.026} & \textbf{0.626$\pm$0.035} & \textbf{0.924$\pm$0.008} \\ \hline
\end{tabular}
\end{table*}

\section{Experiments}
This section provides the main evaluation results and the corresponding analysis whilst more results are shown in the Supplementary Material.
\subsection{Datasets}

In this study, two image sets were used for the evaluation of the proposed method, namely, the public Camelyon-16 breast cancer dataset \cite{bejnordi2017diagnostic} and a collection of colorectal cancer tissue microarray images collected from a tertiary centre.


 \textbf{Camelyon-16}: was released initially for the task of breast cancer metastasis detection, containing in total 400 whole slide images. In our experiments, 193 images with tumor regions in them were cropped from the 158 positive WSIs of Camelyon16. Each image is with a resolution of 4,096x4,096 at 20X magnification. The tumor regions in each image are the target to segment. 

\textbf{Colorectal tissue cores} The colorectal tissue cores as part of a microarray (TMA), in which the cores  (tumour and adjacent normal) were stained with smooth muscle actin (SMA) and counterstained by haematoxylin were used as part of the evaluation. In total 100 tissue cores of 50 cases from TMAs with detailed annotations were used.   Each image is with a resolution of 768x768 at 10X magnification, and contains one tissue core. The stroma regions in these tissue cores are the target to segment.

\subsection{Evaluation and Implementation}
For each dataset, four-folders cross-validation strategy was adopted for the evaluation. The images in a dataset were randomly split into four folders. In each experiment, one folder of images was used as the test set, and the remaining folders together were used for training, within which the images were further split into training set and validation set with a ratio of 4:1. The model achieved the lowest partial cross entropy on the validation set was saved for evaluation on the test set. Note, for colorectal tissue cores the splitting was based on patient-level.

The mean and standard deviation values of performance metrics including Dice Coefficient,  accuracy, precision, Cohen's Kappa, and area under the curve (AUC) are reported.

As an image from Camelyon-16 is with high resolution, the convolutional component of a ResNet-34 \cite{he2016deep} was used as the feature extractor (the yellow area in Figure.\ref{overview})  to extract a feature map with a dimension of 256x256x256 to forward to the segmentation network which is a U-Net \cite{ronneberger2015u} (the orange area in Figure.\ref{overview}). The images from colorectal tissue core set were directly fed into the segmentation network for segmentation, skipping the feature extraction step. For colorectal tissue core images, each pixel annotation is expanded into a 3x3 region in the corresponding partial annotated mask. 

The annotated points were randomly generated according to the ground-truth labels and were fixed for all the experiments. Without particular specification, three pairs of in-target and out-of-target points were used for the Camelyon-16 dataset, while for the colorectal tissue core set, this number was five. 

For training, an Adam optimizer \cite{kingma2014adam} was used for each model with a weight decay of 0.0001 and the learning rate was 0.0001. 30 epochs were used for training the model on the Camelyon-16 dataset, and 50 for that of the colorectal tissue core set.

\subsection{Comparisons with related methods}

The performance of the proposed method is compared with the fully-supervised method, the baseline method (partial cross-entropy), and existing representative methods, including MixMatch-based semi-supervised \cite{gao2020renal}, uncertainty-aware (UA) mean teacher \cite{liu2022weakly}, conditional random field (CRF) loss \cite{tang2018regularized}, dual branch ensemble \cite{luo2022scribble}, and the method of Peng \etal \cite{peng2019semi} which utilizes cosine similarity maps. In addition to the pre-existing deep learning methods, we also included the performances of the two following methods: (1) `Direct Variational Regularization' directly applies the edge-aware variational model (Eq.\ref{eq_edge_cv}) on the segmentation output of the deep segmentation model and the corresponding result is used to supervise the training of the deep segmentation model; (2) `Supervised By Contrast Map' directly uses the mean contrast map to provide extra supervision for the deep segmentation model. 

The comparison results are shown in Table.\ref{tab_compare_otherMethod}. Without surprise, the fully-supervised method achieves the best performance on both datasets with the highest values of all the performance metrics, which are followed by the proposed method. The other existing methods for comparison are significantly inferior to the fully-supervised method and ours. 
Particularly for the Dice Coefficient, the proposed method is at least 10\% higher than other existing methods, and is only 6.5\% lower than the fully-supervised on the Camelyon16 dataset. On the colorectal tissue core dataset, the proposed method is at least 6\% higher than other pre-existing methods in Dice coefficient, and is about 6\% lower than the fully-supervised method. Notably, the `Supervised by Contrast Map' achieves promising performances that are close to the proposed method, indicating that the proposed contrast maps can already provide good quality of segmentation, yet the proposed method which applies the variational model on the contrast maps can further improve the performances. Besides, although the `direct variational regularization' method outperforms most pre-existing methods, it is still not comparable with the proposed method.

\begin{figure*}[t]
  \centering
   \includegraphics[width=0.9\linewidth]{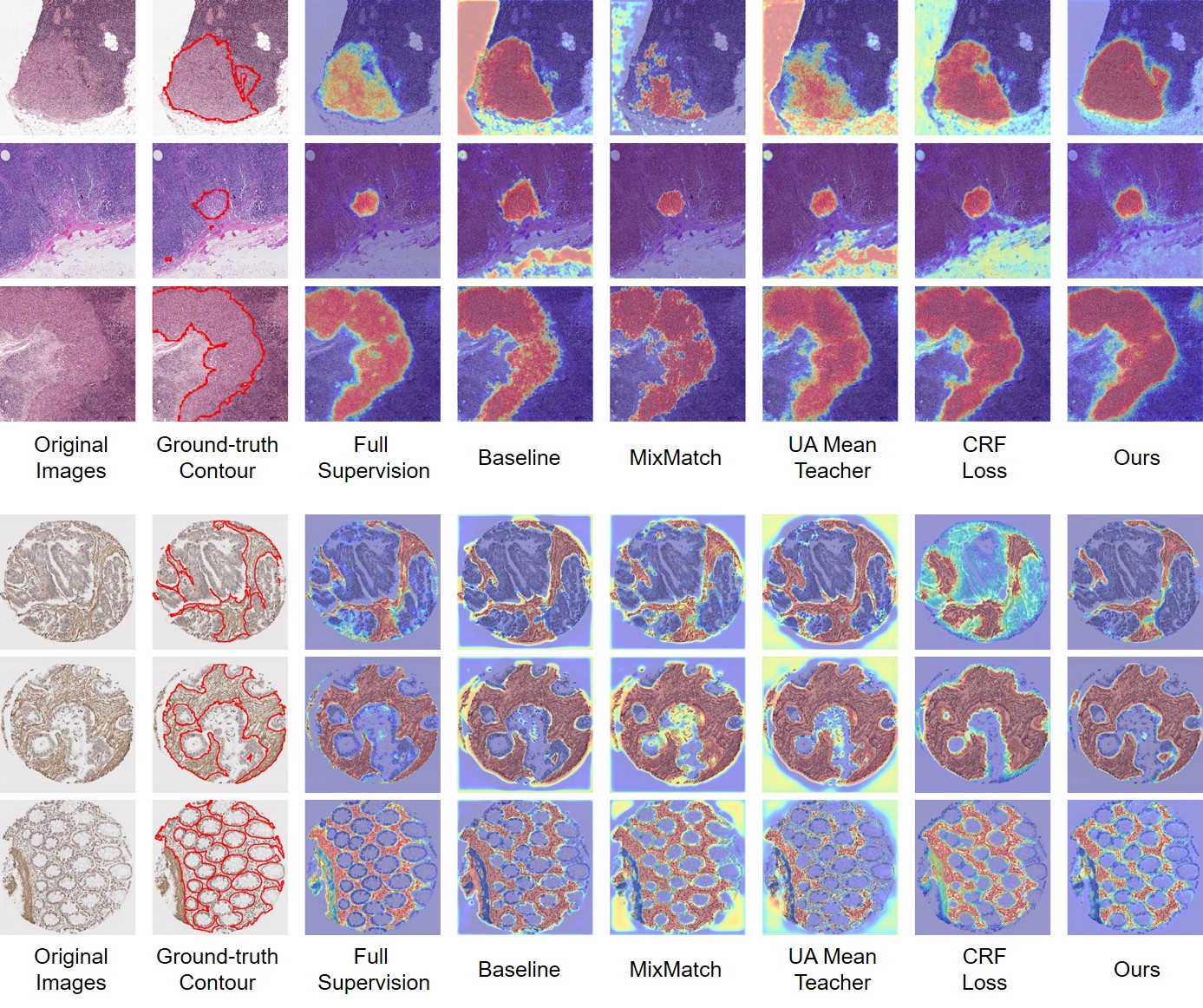}
   \caption{Heatmaps for different methods. Top: examples from the Camelyon-16; Bottom:  examples of the colorectal tissue cores. More are provided in the Supplementary.}
   \label{fig_heatmap}
\end{figure*}

The heatmaps presented in Figure.\ref{fig_heatmap} show some qualitative results. Clearly, the segmentation results from the proposed method are more regionally consistent, usually have smoother boundaries, and are closest to the results of the fully-supervision than those of other methods. It is noticeable that some methods have the issue of wrongly giving some neutral regions comparably strong activations. These regions, such as the sparse tissue regions, do not have a large proportion in most images, therefore the number of random annotated points on them is very few. As a result, there is no sufficient supervision information for a model to learn to recognize them as non-target regions. This problem is particularly severe with the UA mean teacher method \cite{liu2022weakly}. It is probably because the update of the teacher model is lagging behind the student model, yet the wrong recognition of the neutral region from the slowly updated teacher serves to supervise the student model, which will enhance the errors. Apart from the fully-supervised method, the CRF-Loss method \cite{tang2018regularized} and the proposed method are the two that are robust to the issue of insufficient labelling of un-target regions, and both methods consider the regional correlations, yet the CRF-Loss method tends to overly expand the segmentation.

\subsection{Ablation Study}
The ablation study was conducted on the publicly available dataset Camelyon-16 for repeatability. 

\begin{figure}[h]
     \centering
     \begin{subfigure}[b]{0.4\textwidth}
         \centering
         \includegraphics[width=1\linewidth]{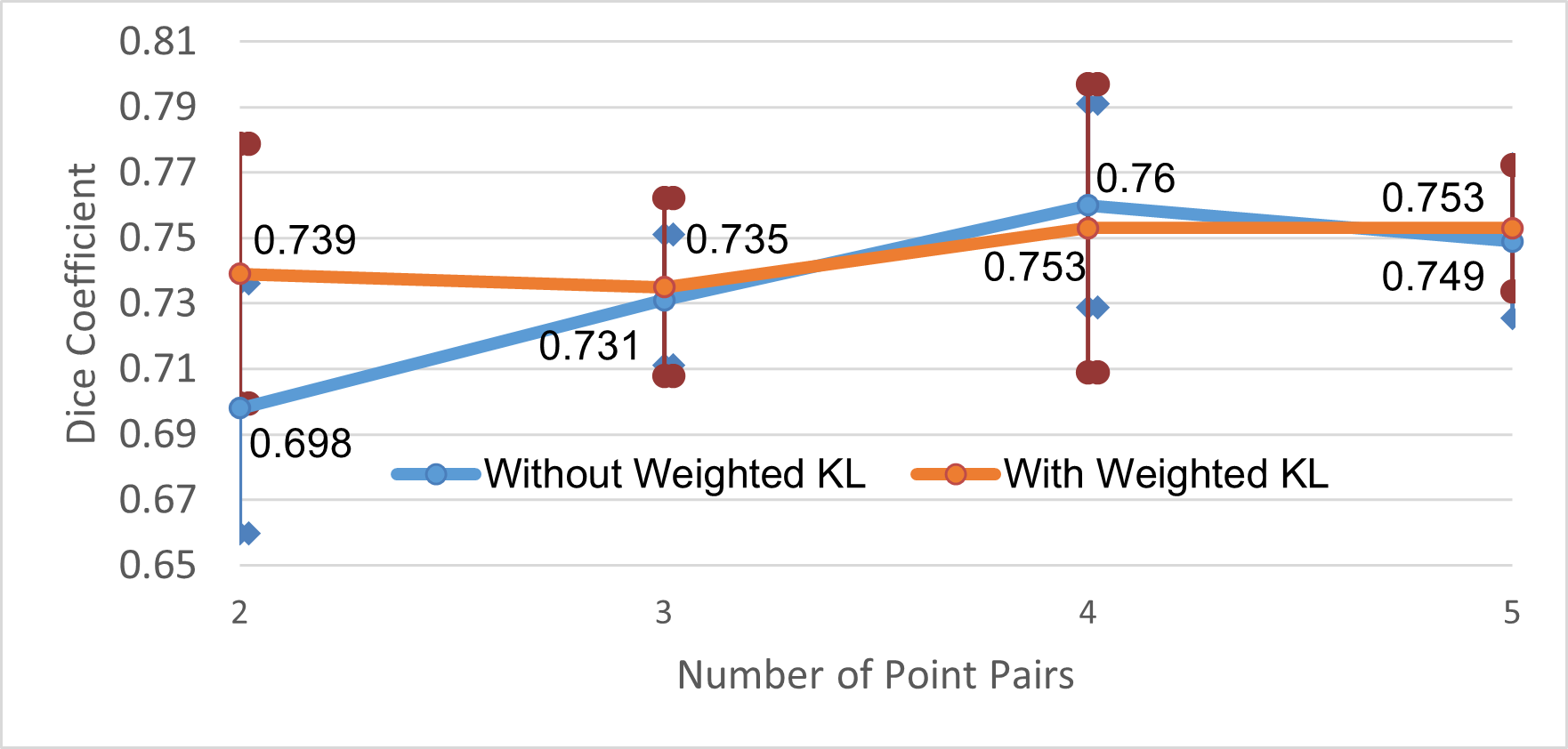}
         \caption{ Dice coefficient values for different numbers of in-target and out-of-target annotated points. }
         \label{sfig_numpoint}
     \end{subfigure}
     \hfill
     \begin{subfigure}[b]{0.4\textwidth}
         \centering
         \includegraphics[width=1\linewidth]{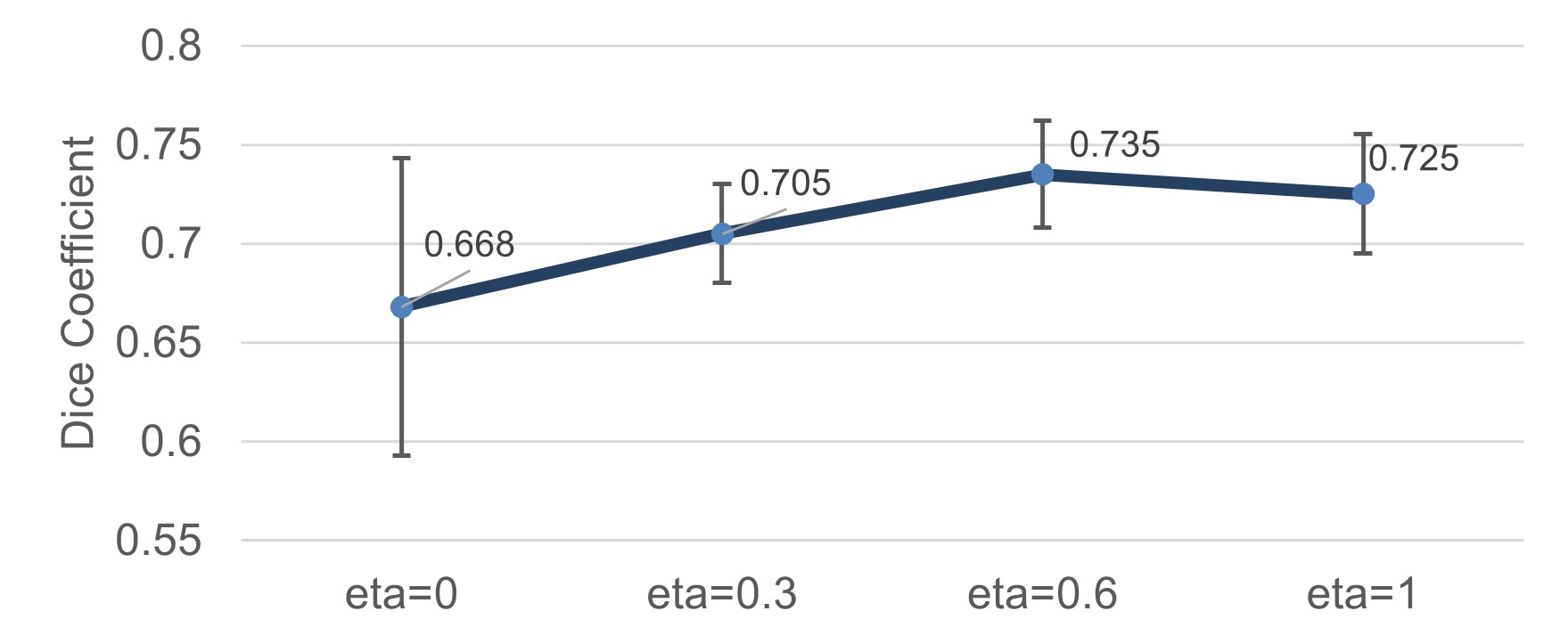}
         \caption{Dice coefficient values for different values of $\eta$}
         \label{sfig_eta}
     \end{subfigure}
        \caption{Performances for different configurations on Camelyon-16.}
        \label{fig_ablation}
\end{figure}

Figure.\ref{sfig_numpoint} shows the Dice coefficient values with and without the weighted KL divergence supervision when using different numbers of pairs of in-target and out-of-target annotated points. The case without weighted KL means the obtained contrast-based variational segmentations were cut-off to binary masks and then used as the pseudo labels for the cross-entropy minimization. As expected, more numbers of pairs involved in training tend to result in better performances. We can also see that when the adopted numbers are equal to or larger than three, the weighted KL loss function shows no obvious superiority over the cross-entropy loss function. However, when the pair numbers are only two, the weighted KL loss function performs significantly better than the counter-part, and is not very inferior to those with more annotated pairs. This phenomenon suggests the robustness of the weighted KL to weaker supervision information. 


Figure.\ref{sfig_eta} shows the effect of $\eta$, which is the weight used in the contrast map calculation (Eq.(\ref{eq_contrast})). The curve shows when $\eta=0.6$, the model achieves the highest Dice coefficient value of 0.735. Particularly, when $\eta =0$ which means no out-of-target points are involved in calculating the contrast maps, the corresponding Dice coefficient is merely 0.668, the lowest value among the four conditions.

\section{Limitation}
The proposed contrast-based variational model sometimes is not able to deal with some complicated cases. Besides, at present there is only CPU version for the solver to optimize the energy functional, thus the time consumption for calculation is comparably considerable. Please refer to the Supplementary for more details.

\section{Conclusion}

We propose an end-to-end method for histopathology image segmentation, with weak supervision information from a few points annotations. The proposed method benefits from the proposed contrast-based variational model (CVM), which provides reliable segmentations as the complementary supervision information, in addition to the weak point supervision, to train the deep segmentation model. The experimental results show a clear superiority of the proposed method to existing methods. It is demonstrated that the trained model is able to generate segmentation results with more consistent regions and smoother boundaries, and is more robust to unlabeled novel regions in comparison to existing methods. We expect the proposed method can be served as a useful tool and can substantially alleviate the burden of the labour-intense annotation for histopathology images. We also expect the proposed method to be applicable to other image modalities if the learnt features are distinct w.r.t. categories.

\section{Acknowledgement}

This work was partly supported by a UK EPSRC GCRF/Newton Consolidation Awards (grant number: EP/X527683/1), and UK EPSRC (grant number: EP/W522399/1).



\clearpage

{\small
\bibliographystyle{ieee_fullname}
\bibliography{egbib}

\begin{thebibliography}{10}\itemsep=-1pt

\bibitem{ambrosio1990approximation}
Luigi Ambrosio and Vincenzo~Maria Tortorelli.
\newblock Approximation of functional depending on jumps by elliptic functional
  via t-convergence.
\newblock {\em Communications on Pure and Applied Mathematics},
  43(8):999--1036, 1990.

\bibitem{bejnordi2017diagnostic}
Babak~Ehteshami Bejnordi, Mitko Veta, Paul~Johannes Van~Diest, Bram
  Van~Ginneken, Nico Karssemeijer, Geert Litjens, Jeroen~AWM Van Der~Laak,
  Meyke Hermsen, Quirine~F Manson, Maschenka Balkenhol, et~al.
\newblock Diagnostic assessment of deep learning algorithms for detection of
  lymph node metastases in women with breast cancer.
\newblock {\em JAMA}, 318(22):2199--2210, 2017.

\bibitem{berthelot2019mixmatch}
David Berthelot, Nicholas Carlini, Ian Goodfellow, Nicolas Papernot, Avital
  Oliver, and Colin~A Raffel.
\newblock Mixmatch: A holistic approach to semi-supervised learning.
\newblock {\em Advances in Neural Information Processing Systems}, 32, 2019.

\bibitem{boykov2001interactive}
Yuri~Y Boykov and M-P Jolly.
\newblock Interactive graph cuts for optimal boundary \& region segmentation of
  objects in nd images.
\newblock In {\em Proceedings eighth IEEE International Conference on Computer
  Vision. ICCV 2001}, volume~1, pages 105--112. IEEE, 2001.

\bibitem{bresson2007fast}
Xavier Bresson, Selim Esedoḡlu, Pierre Vandergheynst, Jean-Philippe Thiran,
  and Stanley Osher.
\newblock Fast global minimization of the active contour/snake model.
\newblock {\em Journal of Mathematical Imaging and Vision}, 28(2):151--167,
  2007.

\bibitem{burrows2020reproducible}
Liam Burrows, Weihong Guo, Ke Chen, and Francesco Torella.
\newblock Reproducible kernel hilbert space based global and local image
  segmentation.
\newblock {\em Inverse Problems \& Imaging}, 2020.

\bibitem{cai2013two}
Xiaohao Cai, Raymond Chan, and Tieyong Zeng.
\newblock A two-stage image segmentation method using a convex variant of the
  mumford--shah model and thresholding.
\newblock {\em SIAM Journal on Imaging Sciences}, 6(1):368--390, 2013.

\bibitem{can2018learning}
Yigit~B Can, Krishna Chaitanya, Basil Mustafa, Lisa~M Koch, Ender Konukoglu,
  and Christian~F Baumgartner.
\newblock Learning to segment medical images with scribble-supervision alone.
\newblock In {\em Deep Learning in Medical Image Analysis and Multimodal
  Learning for Clinical Decision Support}, pages 236--244. Springer, 2018.

\bibitem{chambolle1999finite}
Antonin Chambolle.
\newblock Finite-differences discretizations of the mumford-shah functional.
\newblock {\em ESAIM: Mathematical Modelling and Numerical Analysis},
  33(2):261--288, 1999.

\bibitem{chan2006algorithms}
Tony~F Chan, Selim Esedoglu, and Mila Nikolova.
\newblock Algorithms for finding global minimizers of image segmentation and
  denoising models.
\newblock {\em SIAM Journal on Applied Mathematics}, 66(5):1632--1648, 2006.

\bibitem{chan2001active}
Tony~F Chan and Luminita~A Vese.
\newblock Active contours without edges.
\newblock {\em IEEE Transactions on Image Processing}, 10(2):266--277, 2001.

\bibitem{cheng2020self}
Hsien-Tzu Cheng, Chun-Fu Yeh, Po-Chen Kuo, Andy Wei, Keng-Chi Liu, Mong-Chi Ko,
  Kuan-Hua Chao, Yu-Ching Peng, and Tyng-Luh Liu.
\newblock Self-similarity student for partial label histopathology image
  segmentation.
\newblock In {\em European Conference on Computer Vision}, pages 117--132.
  Springer, 2020.

\bibitem{coupland2023application}
Sarah~E Coupland, Hongrun Zhang, Hayley Jones, and Yalin Zheng.
\newblock Application of deep learning models in the evaluation of
  histopathology of uveal melanoma.
\newblock In {\em Global Perspectives in Ocular Oncology}, pages 211--216.
  Springer, 2023.

\bibitem{dosovitskiy2020image}
Alexey Dosovitskiy, Lucas Beyer, Alexander Kolesnikov, Dirk Weissenborn,
  Xiaohua Zhai, Thomas Unterthiner, Mostafa Dehghani, Matthias Minderer, Georg
  Heigold, Sylvain Gelly, et~al.
\newblock An image is worth 16x16 words: Transformers for image recognition at
  scale.
\newblock {\em arXiv preprint arXiv:2010.11929}, 2020.

\bibitem{gao2020renal}
Zeyu Gao, Pargorn Puttapirat, Jiangbo Shi, and Chen Li.
\newblock Renal cell carcinoma detection and subtyping with minimal point-based
  annotation in whole-slide images.
\newblock In {\em International Conference on Medical Image Computing and
  Computer-Assisted Intervention}, pages 439--448. Springer, 2020.

\bibitem{gout2005segmentation}
Christian Gout, Carole Le~Guyader, and Luminita Vese.
\newblock Segmentation under geometrical conditions using geodesic active
  contours and interpolation using level set methods.
\newblock {\em Numerical Algorithms}, 39(1):155--173, 2005.

\bibitem{grady2006random}
Leo Grady.
\newblock Random walks for image segmentation.
\newblock {\em IEEE Transactions on Pattern Analysis and Machine Intelligence},
  28(11):1768--1783, 2006.

\bibitem{he2016deep}
Kaiming He, Xiangyu Zhang, Shaoqing Ren, and Jian Sun.
\newblock Deep residual learning for image recognition.
\newblock In {\em Proceedings of the IEEE Conference on Computer Vision and
  Pattern Recognition}, pages 770--778, 2016.

\bibitem{kervadec2019constrained}
Hoel Kervadec, Jose Dolz, Meng Tang, Eric Granger, Yuri Boykov, and Ismail~Ben
  Ayed.
\newblock Constrained-cnn losses for weakly supervised segmentation.
\newblock {\em Medical Image Analysis}, 54:88--99, 2019.

\bibitem{kim2019mumford}
Boah Kim and Jong~Chul Ye.
\newblock Mumford--shah loss functional for image segmentation with deep
  learning.
\newblock {\em IEEE Transactions on Image Processing}, 29:1856--1866, 2019.

\bibitem{kingma2014adam}
Diederik~P Kingma and Jimmy Ba.
\newblock Adam: A method for stochastic optimization.
\newblock {\em arXiv preprint arXiv:1412.6980}, 2014.

\bibitem{lee2020scribble2label}
Hyeonsoo Lee and Won-Ki Jeong.
\newblock Scribble2label: Scribble-supervised cell segmentation via
  self-generating pseudo-labels with consistency.
\newblock In {\em International Conference on Medical Image Computing and
  Computer-Assisted Intervention}, pages 14--23. Springer, 2020.

\bibitem{li2021dual}
Bin Li, Yin Li, and Kevin~W Eliceiri.
\newblock Dual-stream multiple instance learning network for whole slide image
  classification with self-supervised contrastive learning.
\newblock In {\em Proceedings of the IEEE/CVF Conference on Computer Vision and
  Pattern Recognition}, pages 14318--14328, 2021.

\bibitem{li2019weakly}
Chao Li, Xinggang Wang, Wenyu Liu, Longin~Jan Latecki, Bo Wang, and Junzhou
  Huang.
\newblock Weakly supervised mitosis detection in breast histopathology images
  using concentric loss.
\newblock {\em Medical Image Analysis}, 53:165--178, 2019.

\bibitem{li2023task}
Honglin Li, Chenglu Zhu, Yunlong Zhang, Yuxuan Sun, Zhongyi Shui, Wenwei Kuang,
  Sunyi Zheng, and Lin Yang.
\newblock Task-specific fine-tuning via variational information bottleneck for
  weakly-supervised pathology whole slide image classification.
\newblock {\em arXiv preprint arXiv:2303.08446}, 2023.

\bibitem{lin2016scribblesup}
Di Lin, Jifeng Dai, Jiaya Jia, Kaiming He, and Jian Sun.
\newblock Scribblesup: Scribble-supervised convolutional networks for semantic
  segmentation.
\newblock In {\em Proceedings of the IEEE Conference on Computer Vision and
  Pattern Recognition}, pages 3159--3167, 2016.

\bibitem{lin2023interventional}
Tiancheng Lin, Zhimiao Yu, Hongyu Hu, Yi Xu, and Chang~Wen Chen.
\newblock Interventional bag multi-instance learning on whole-slide
  pathological images.
\newblock {\em arXiv preprint arXiv:2303.06873}, 2023.

\bibitem{litjens2016deep}
Geert Litjens, Clara~I S{\'a}nchez, Nadya Timofeeva, Meyke Hermsen, Iris
  Nagtegaal, Iringo Kovacs, Christina Hulsbergen-Van De~Kaa, Peter Bult, Bram
  Van~Ginneken, and Jeroen Van Der~Laak.
\newblock Deep learning as a tool for increased accuracy and efficiency of
  histopathological diagnosis.
\newblock {\em Scientific Reports}, 6(1):1--11, 2016.

\bibitem{liu2018weighted}
Chunxiao Liu, Michael Kwok-Po Ng, and Tieyong Zeng.
\newblock Weighted variational model for selective image segmentation with
  application to medical images.
\newblock {\em Pattern Recognition}, 76:367--379, 2018.

\bibitem{liu2022weakly}
Xiaoming Liu, Quan Yuan, Yaozong Gao, Kelei He, Shuo Wang, Xiao Tang, Jinshan
  Tang, and Dinggang Shen.
\newblock Weakly supervised segmentation of covid19 infection with scribble
  annotation on ct images.
\newblock {\em Pattern Recognition}, 122:108341, 2022.

\bibitem{luo2021semi}
Xiangde Luo, Jieneng Chen, Tao Song, and Guotai Wang.
\newblock Semi-supervised medical image segmentation through dual-task
  consistency.
\newblock In {\em Proceedings of the AAAI Conference on Artificial
  Intelligence}, volume~35, pages 8801--8809, 2021.

\bibitem{luo2022scribble}
Xiangde Luo, Minhao Hu, Wenjun Liao, Shuwei Zhai, Tao Song, Guotai Wang, and
  Shaoting Zhang.
\newblock Scribble-supervised medical image segmentation via dual-branch
  network and dynamically mixed pseudo labels supervision.
\newblock {\em arXiv preprint arXiv:2203.02106}, 2022.

\bibitem{mumford1989optimal}
David~Bryant Mumford and Jayant Shah.
\newblock Optimal approximations by piecewise smooth functions and associated
  variational problems.
\newblock {\em Communications on Pure and Applied Mathematics}, 1989.

\bibitem{peng2019semi}
Liying Peng, Lanfen Lin, Hongjie Hu, Yue Zhang, Huali Li, Yutaro Iwamoto,
  Xian-Hua Han, and Yen-Wei Chen.
\newblock Semi-supervised learning for semantic segmentation of emphysema with
  partial annotations.
\newblock {\em IEEE Journal of Biomedical and Health Informatics},
  24(8):2327--2336, 2019.

\bibitem{pock2009algorithm}
Thomas Pock, Daniel Cremers, Horst Bischof, and Antonin Chambolle.
\newblock An algorithm for minimizing the mumford-shah functional.
\newblock In {\em 2009 IEEE 12th International Conference on Computer Vision},
  pages 1133--1140. IEEE, 2009.

\bibitem{qu2020weakly}
Hui Qu, Pengxiang Wu, Qiaoying Huang, Jingru Yi, Zhennan Yan, Kang Li,
  Gregory~M Riedlinger, Subhajyoti De, Shaoting Zhang, and Dimitris~N Metaxas.
\newblock Weakly supervised deep nuclei segmentation using partial points
  annotation in histopathology images.
\newblock {\em IEEE Transactions on Medical Imaging}, 39(11):3655--3666, 2020.

\bibitem{roberts2019convex}
Michael Roberts, Ke Chen, and Klaus~L Irion.
\newblock A convex geodesic selective model for image segmentation.
\newblock {\em Journal of Mathematical Imaging and Vision}, 61(4):482--503,
  2019.

\bibitem{ronneberger2015u}
Olaf Ronneberger, Philipp Fischer, and Thomas Brox.
\newblock U-net: Convolutional networks for biomedical image segmentation.
\newblock In {\em International Conference on Medical Image Computing and
  Computer-Assisted Intervention}, pages 234--241. Springer, 2015.

\bibitem{rother2004grabcut}
Carsten Rother, Vladimir Kolmogorov, and Andrew Blake.
\newblock " grabcut" interactive foreground extraction using iterated graph
  cuts.
\newblock {\em ACM Transactions on Graphics (TOG)}, 23(3):309--314, 2004.

\bibitem{rudin1992nonlinear}
Leonid~I Rudin, Stanley Osher, and Emad Fatemi.
\newblock Nonlinear total variation based noise removal algorithms.
\newblock {\em Physica D: Nonlinear Phenomena}, 60(1-4):259--268, 1992.

\bibitem{shao2021transmil}
Zhuchen Shao, Hao Bian, Yang Chen, Yifeng Wang, Jian Zhang, Xiangyang Ji,
  et~al.
\newblock Transmil: Transformer based correlated multiple instance learning for
  whole slide image classification.
\newblock {\em Advances in Neural Information Processing Systems},
  34:2136--2147, 2021.

\bibitem{spencer2015convex}
Jack Spencer and Ke Chen.
\newblock A convex and selective variational model for image segmentation.
\newblock {\em Communications in Mathematical Sciences}, 13(6):1453--1472,
  2015.

\bibitem{takahama2019multi}
Shusuke Takahama, Yusuke Kurose, Yusuke Mukuta, Hiroyuki Abe, Masashi Fukayama,
  Akihiko Yoshizawa, Masanobu Kitagawa, and Tatsuya Harada.
\newblock Multi-stage pathological image classification using semantic
  segmentation.
\newblock In {\em Proceedings of the IEEE/CVF International Conference on
  Computer Vision}, pages 10702--10711, 2019.

\bibitem{tang2018normalized}
Meng Tang, Abdelaziz Djelouah, Federico Perazzi, Yuri Boykov, and Christopher
  Schroers.
\newblock Normalized cut loss for weakly-supervised cnn segmentation.
\newblock In {\em Proceedings of the IEEE Conference on Computer Vision and
  Pattern Recognition}, pages 1818--1827, 2018.

\bibitem{tang2018regularized}
Meng Tang, Federico Perazzi, Abdelaziz Djelouah, Ismail Ben~Ayed, Christopher
  Schroers, and Yuri Boykov.
\newblock On regularized losses for weakly-supervised cnn segmentation.
\newblock In {\em Proceedings of the European Conference on Computer Vision
  (ECCV)}, pages 507--522, 2018.

\bibitem{valvano2021learning}
Gabriele Valvano, Andrea Leo, and Sotirios~A Tsaftaris.
\newblock Learning to segment from scribbles using multi-scale adversarial
  attention gates.
\newblock {\em IEEE Transactions on Medical Imaging}, 40(8):1990--2001, 2021.

\bibitem{weickert1998efficient}
Joachim Weickert, BM~Ter~Haar Romeny, and Max~A Viergever.
\newblock Efficient and reliable schemes for nonlinear diffusion filtering.
\newblock {\em IEEE Transactions on Image Processing}, 7(3):398--410, 1998.

\bibitem{xiang2023exploring}
Jinxi Xiang and Jun Zhang.
\newblock Exploring low-rank property in multiple instance learning for whole
  slide image classification.
\newblock In {\em The Eleventh International Conference on Learning
  Representations}, 2023.

\bibitem{yoo2019pseudoedgenet}
Inwan Yoo, Donggeun Yoo, and Kyunghyun Paeng.
\newblock Pseudoedgenet: Nuclei segmentation only with point annotations.
\newblock In {\em International Conference on Medical Image Computing and
  Computer-Assisted Intervention}, pages 731--739. Springer, 2019.

\bibitem{zhang2020piloting}
Hongrun Zhang, Helen Kalirai, Amelia Acha-Sagredo, Xiaoyun Yang, Yalin Zheng,
  and Sarah~E Coupland.
\newblock Piloting a deep learning model for predicting nuclear bap1
  immunohistochemical expression of uveal melanoma from hematoxylin-and-eosin
  sections.
\newblock {\em Translational Vision Science \& Technology}, 9(2):50--50, 2020.

\bibitem{zhang2022dtfd}
Hongrun Zhang, Yanda Meng, Yitian Zhao, Yihong Qiao, Xiaoyun Yang, Sarah~E
  Coupland, and Yalin Zheng.
\newblock Dtfd-mil: Double-tier feature distillation multiple instance learning
  for histopathology whole slide image classification.
\newblock In {\em Proceedings of the IEEE/CVF Conference on Computer Vision and
  Pattern Recognition}, pages 18802--18812, 2022.

\bibitem{zhao2020weakly}
Tianyi Zhao and Zhaozheng Yin.
\newblock Weakly supervised cell segmentation by point annotation.
\newblock {\em IEEE Transactions on Medical Imaging}, 40(10):2736--2747, 2020.

\end{thebibliography}
}

\end{document}